\documentclass[8.5pt,twoside,twocolumn]{article}
\usepackage[super,sort&compress,comma]{natbib} 
\usepackage{mhchem}
\usepackage{times,mathptmx}
\usepackage{sectsty}
\usepackage{balance} 

\usepackage{graphicx} %eps figures can be used instead
\usepackage{lastpage}
\usepackage[format=plain,justification=raggedright,singlelinecheck=false,font=small,labelfont=bf,labelsep=space]{caption} 
\usepackage{fancyhdr}
\begin{document}

\thispagestyle{plain}
\fancypagestyle{plain}{
\fancyhead[L]{\includegraphics[height=8pt]{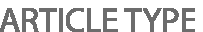}}
\fancyhead[C]{\hspace{-1cm}\includegraphics[height=20pt]{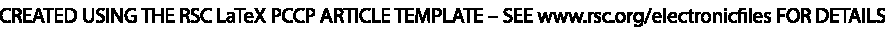}}
\fancyhead[R]{\includegraphics[height=10pt]{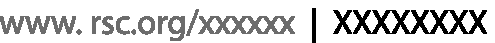}\vspace{-0.2cm}}
\renewcommand{\headrulewidth}{1pt}}
\renewcommand{\thefootnote}{\fnsymbol{footnote}}
\renewcommand\footnoterule{\vspace*{1pt}% 
\hrule width 3.4in height 0.4pt \vspace*{5pt}} 
\setcounter{secnumdepth}{5}

\makeatletter 
\def\subsubsection{\@startsection{subsubsection}{3}{10pt}{-1.25ex plus -1ex minus -.1ex}{0ex plus 0ex}{\normalsize\bf}} 
\def\paragraph{\@startsection{paragraph}{4}{10pt}{-1.25ex plus -1ex minus -.1ex}{0ex plus 0ex}{\normalsize\textit}} 
\renewcommand\@biblabel[1]{#1}            
\renewcommand\@makefntext[1]% 
{\noindent\makebox[0pt][r]{\@thefnmark\,}#1}
\makeatother 
\renewcommand{\figurename}{\small{Fig.}~}
\sectionfont{\large}
\subsectionfont{\normalsize} 

\fancyfoot{}
\fancyfoot[LO,RE]{\vspace{-7pt}\includegraphics[height=9pt]{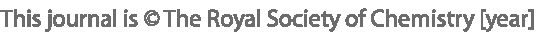}}
\fancyfoot[CO]{\vspace{-7.2pt}\hspace{12.2cm}\includegraphics{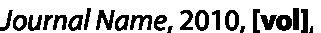}}
\fancyfoot[CE]{\vspace{-7.5pt}\hspace{-13.5cm}\includegraphics{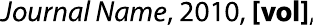}}
\fancyfoot[RO]{\footnotesize{\sffamily{1--\pageref{LastPage} ~\textbar  \hspace{2pt}\thepage}}}
\fancyfoot[LE]{\footnotesize{\sffamily{\thepage~\textbar\hspace{3.45cm} 1--\pageref{LastPage}}}}
\fancyhead{}
\renewcommand{\headrulewidth}{1pt} 
\renewcommand{\footrulewidth}{1pt}
\setlength{\arrayrulewidth}{1pt}
\setlength{\columnsep}{6.5mm}
\setlength\bibsep{1pt}

\twocolumn[
  \begin{@twocolumnfalse}
\noindent\LARGE{\textbf{Phase behavior of binary mixtures of hard convex polyhedra$^\dag$}}
\vspace{0.6cm}

\noindent\large{\textbf{Mihir R. Khadilkar\textit{$^{a}$}, Umang Agarwal \textit{$^{b}$}and
Fernando A. Escobedo $^{\ast}$\textit{$^{c\ddag}$}}}\vspace{0.5cm}
%Please note that \ast indicates the corresponding author(s) but no footnote text is required. 

\noindent\textit{\small{\textbf{Received Xth XXXXXXXXXX 20XX, Accepted Xth XXXXXXXXX 20XX\newline
First published on the web Xth XXXXXXXXXX 200X}}}

\noindent \textbf{\small{DOI: 10.1039/b000000x}}
\vspace{0.6cm}
%Please do not change this text.

\noindent \normalsize{ 
Shape anisotropy of colloidal nanoparticles has emerged as an important design variable for engineering assemblies with targeted structure and properties. In particular, a number of polyhedral nanoparticles have been shown to exhibit a  rich phase behavior [Agarwal  \textit{et al., Nature Materials}, 2011, \textbf{10}, 230]. Since real synthesized particles have polydispersity not only in size but also in shape, we explore here the phase behavior of binary mixtures of hard convex polyhedra having similar sizes but different shapes. Choosing representative particle shapes from those readily synthesizable, we study in particular four mixtures: (i) cubes and spheres (with spheres providing a non-polyhedral reference case), (ii) cubes and truncated octahedra, (iii) cubes and cuboctahedra, and (iv) cuboctahedra and truncated octahedra. The phase behavior of  such mixtures is dependent on the interplay of mixing and packing entropy, which can give rise to  miscible or  phase-separated states. The extent of mixing of two such particle types is expected to depend on  the degree of shape similarity, relative sizes, composition, and compatibility of the crystal structures formed by the pure components. While expectedly the binary systems studied  exhibit phase separation at high pressures due to the incompatible pure-component crystal structures, our study shows that the essential qualitative trends in miscibility and phase separation can be correlated to properties of the pure components, such as the relative values of the order-disorder transition pressure (ODP) of each component. Specifically, if for a mixture A+B we have that ODP$_B<$ODP$_A$ and $\Delta$ ODP = ODP$_A$ - ODP$_B$, then at any particular pressure where phase separation occurs, the larger the $\Delta$ ODP the lower the solubility of A in the B-rich ordered phase and the higher the solubility of B in the A-rich ordered phase.}
\vspace{0.5cm}
 \end{@twocolumnfalse}
  ]

\tableofcontents
\section{Introduction}
%Footnotes
%\footnotetext{\dag~Electronic Supplementary Information (ESI) available: [details of any supplementary information available should be included here]. See DOI: 10.1039/b000000x/}

%Please use \dag to cite the ESI in the main text of the article.
%If you article does not have ESI please remove the the \dag symbol from the title and the above footnotetext.
\footnotetext{\textit{$^{a}$~Department of Physics, Cornell University, Ithaca, New York. }}
\footnotetext{\textit{$^{b}$~School of Chemical and Biomolecular Engineering, Cornell University, Ithaca, New York. }}
\footnotetext{\textit{$^{c}$~School of Chemical and Biomolecular Engineering, Cornell University, Ithaca, New York. E-mail: fe13@cornell.edu}}

%additional addresses can be cited as above using the lower-case letters, c, d, e... If all authors are from the same address, no letter is required

%\footnotetext{\ddag~Additional footnotes to the title and authors can be included \emph{e.g.}\ `Present address:' or `These authors contributed equally to this work' as above using the symbols: \ddag, \textsection, and \P. Please place the appropriate symbol next to the author's name and include a \texttt{\textbackslash footnotetext} entry in the the correct place in the list.}

Interest in material design based on nanocrystal assemblies has been rapidly increasing over the past decade. The choice of building blocks,  size and shape,  composition and surface functionalization offers multiple avenues to taylor the properties of these assemblies. Going beyond the prototypical spherical nanoparticles, researchers have explored the effect of shape anisotropy on colloidal self-assembly afforded by the synthesis of cubes, rods, plates, disks and particles with core-shell structure \cite{Buining1991}\cite{Adams1998}\cite{Sun2002}\cite{Kooij2000}. With the advent of new particle synthesis methods\cite{Personick2011}\cite{Chiu2011}\cite{Quan2010}\cite{Li2011} polyhedral nanoparticles are becoming ubiquitous in providing varying extent	 of shape `anisotropy'.  Suspensions of a range of polyhedral shapes have been shown to produce interesting partially ordered `mesophases' \cite{Agarwal2011}  \cite{Damasceno2012}\cite{Haji-akbari2012}\cite{John2008} at intermediate volume fractions.  The mixing of nanoparticles of different chemical composition could provide a simple way to prepare ordered structures with more desirable properties; e.g., if a dopant component adds some functionality to the host component in a matrix. If the particles differ not only in composition but also in shape, one may gain additional control over the extent of infiltration by the dopant component, how their particles orient and what spatial environment they encounter. 
\par
\begin{figure}
 \centering
   \includegraphics[height=4cm]{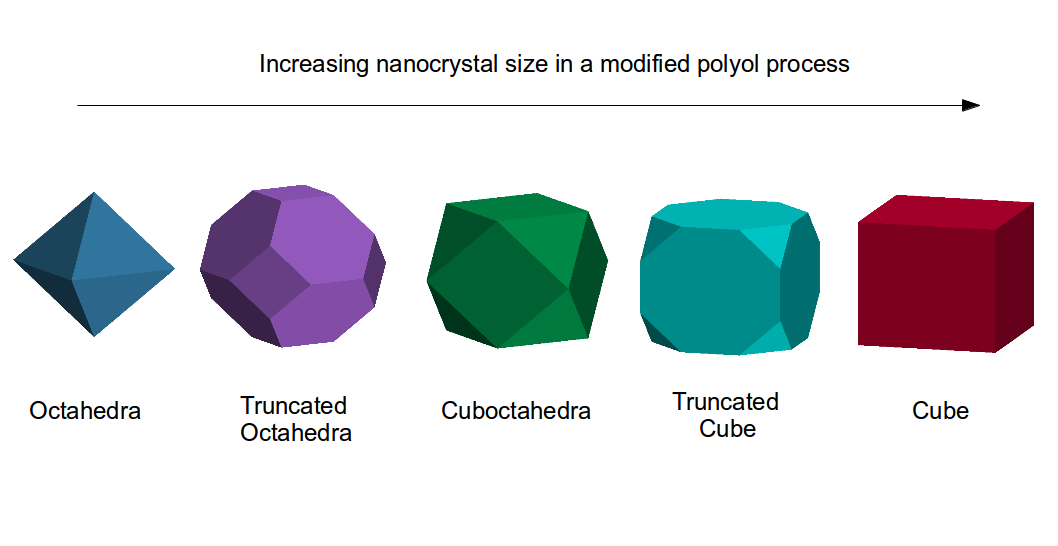}
   \caption{Figure showing some of the polyhedra synthesized at different steps of a modified polyol process. The arrow shows the order of the shapes produced as the reaction is allowed for longer times.}
 \label{fgr:process}
 \end{figure}

A range of different polyhedral nanocrystals have become readily synthesizable by controlling the growth step in a modified polyol process for the formation of  nanocrystals made of gold \cite{Seo2006}\cite{Compton2007} and other inorganic materials\cite{Niu2010}\cite{Long2011} ; Figure \ref{fgr:process} shows some of the particle shapes typically produced this way. This approach also readily leads to  mixtures of these polyhedra (shape bidispersity) by stopping the growth process at different stages. The phase behavior of such binary mixtures (of particles having similar size but different shape) is expected to depend on multiple factors, including  the phase behavior of the individual polyhedra and in particular their order-disorder transition pressure (ODP) and the crystal lattice they assemble into. A specific class of tessellating mixtures\cite{Khadilkar2012} has been studied before; in that case the pure components exhibit incompatible crystal lattices, but at a precise (stoichiometric) composition and size ratio, they can assemble into binary crystalline compounds. However, most mixtures of polyhedra with incompatible crystal lattices (for the monodisperse systems) would be expected  to phase separate at higher volume fractions into ordered phases resembling those of the pure components.

\par
When a mixture of two polyhedral shapes A + B does phase separate, it is important to explore the extent to which they mix in the different phases; i.e., how much B is incorporated in the A-rich phase and how much A is incorporated in the B-rich phase. Such information will be useful, e.g., in predicting the limiting compositions at which such bidisperse mixtures can exist without phase separation. Two components that individually form the same type of mesophase may give rise to mixtures that exhibit a similar mesophase. For example, appreciable mixing of components has been observed in the case of polydisperse mixtures of rigid rods of different length which form  a nematic phase\cite{Escobedo2003}, and of cubes of different size that form a cubatic mesophase\cite{Agarwal2012}. 
\par 
With these considerations in mind, we study here the self-assembly of binary mixtures of convex polyehdra. As a reference case, we study first the cube + sphere mixture (CS), taking the sphere as the infinite-facet limit of a regular polyhedron. Further, we choose three representative binary mixtures, each made from polyhedra that have been synthesized by various experimental methods  \cite{Seo2006}\cite{Compton2007}\cite{Henzie2011} : cubes + truncated octahedra (CTO), cuboctahedra +cubes (COC), and cuboctahedra + truncated octahedra (COTO).  These mixtures represent different degrees  of similarity in terms of rotational symmetry, asphericity \cite{Agarwal2011} and the crystal lattice they individually assemble into. The presence of similar mesophases\cite{Agarwal2011} may also affect this degree of mixing. Large difference in ODPs between components is associated with larger difference in sizes and hence is expected to result in more `incompatible' mixtures, that readily phase separate. Conversely,  components with similar ODPs and  similar mesophases could be  expected to better mix with each other. Our aim in this paper is to determine approximate pressure-composition phase diagrams for each of these mixtures to try to elucidate the relation between the phase behavior of the mixture and that of the pure components. Although there are many experimental methods prevalent in the literature that allow synthesis of concave shapes, as a first step, we aim to keep our analysis simple by restricting ourselves to more common convex shapes.

\par The CS system  represents a limiting case when one of the components (spheres)  has no anisotropy. In this case, the individual shapes and their crystal structures are quite incompatible and they are expected to phase separate above the components' ODPs. In our simulations, we have assumed that the side of the cube is equal to the diameter of the sphere. While such a choice does fulfill the criterion of having components similar in size, it will necessarily affect the precise location of phase boundaries in the phase diagram and also create an asymmetry in the relative diffusivities of the components in a given phase (with the smaller component typically having larger mobility). 
 \begin{figure}
 \label{fgr:systems}
 \centering
   \includegraphics[height=8cm]{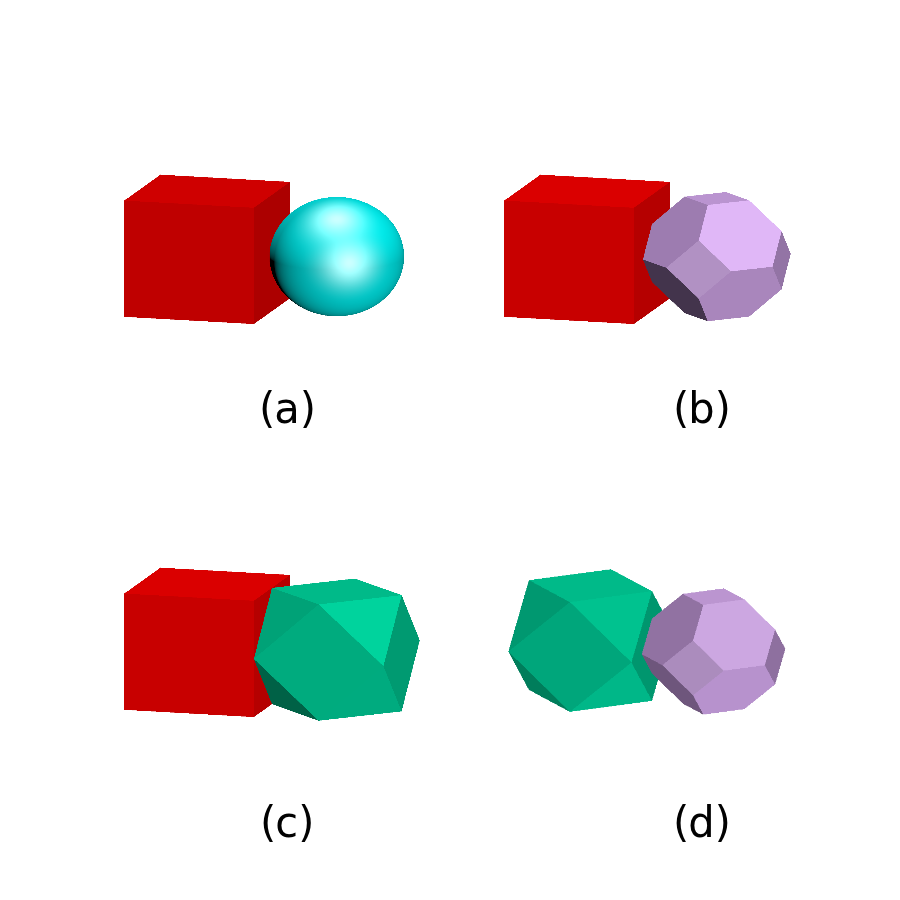}
   \caption{Snapshots showing the 4 mixtures studied: a) CS mixture (cubes+ spheres), b) CTO mixture (cubes+ truncated octehedra), c) COC mixture (cubes + cuboctahedra), d) COTO mixture (cuboctahedra + truncated octahedra). The relative sizes accurately describe the size ratios in simulations.}
  \label{fgr:systems}
 \end{figure}

\par The other mixtures studied, namely, CTO, COC and COTO represent cases of components with varying degrees of similarity in shape and size. The CTO mixture corresponds to components which are rather distant  in the polyol process (see Figure \ref{fgr:process}); this mixture also embodies a peculiar feature where the two individual components are space tessellating (at close packing) but a mixture of any composition is not. Such a feature and the fact that their ordered phases have very different symmetry suggest that the CTO system should be largely incompatible (not unlike the CS system). Considering that cubes and truncated cubes have almost identical phase behavior and very similar shapes, one can consider the mixture of cubes and cuboctahedra (COC) as representing shapes which are essentially neighbors in the polyol process (see Figure \ref{fgr:process}). Further, the partial similarity of the ordered structures of the pure components suggest that the COC mixture  could have an intermediate degree of compatibility. The mixture of cuboctahedra and truncated octahedral (COTO) can be considered as potentially the most compatible given that the components are proximal in the polyol process (Fig. 1) and that both form a rotator mesophase. In each of the mixtures, the size ratios of the components (see next Section) are consistent with those obtained in a typical polyol process\cite{Seo2006}\cite{Compton2007}. \section{\label{sec:methods}{Methodology\protect  }}
In our simulations, we assume the particles to be interacting via hard-core potentials, which amounts to ensuring that particles never overlap. Extensive expansion and compression Monte Carlo (MC) runs were performed to map out the equation of state of each of the mixtures, at constant pressure and particle number (NPT ensemble). Although we observe hysteresis between expansion and compression runs,  we use expansion runs to estimate the transition pressures because they typically need to surmount a smaller free energy barrier at the transition points and are hence expected  to more closely follow thermodynamic behavior.  Unless otherwise indicated, the  mixtures were  simulated at the same composition of 50 \%. We used a total of 2560 particles for CS, CTO and COTO mixtures and 2048 particles for COC mixture. 
As indicated earlier, the size of the components was chosen based on typical results of the modified polyol process. For such a process, Seo et al. \cite{Seo2006} report the average edge lengths of different shapes to be 98, 120 and 145 nm for truncated octahedra, cuboctahedra and cubes respectively. Of course, synthesis conditions can be altered so that a given particle shape can be obtained with different sizes, but choosing sizes consistent with those attainable within a single growth experiment underscores a simple possible strategy to obtain the mixtures studied in this work. 
\par The dimensionless osmotic pressure is defined as 
\begin{equation}
P^* = \frac{Pl^3}{\epsilon},
\end{equation}
where $l = 1$, is the characteristic length in each of the mixtures simulated here. In mixtures involving the cubic particles, this characteristic length is the length of the side of the cube, while in COTO mixture, it is the side of the imaginary cube from which the cuboctahedron is cut.
$\epsilon$ is an arbitrary energy  parameter (set to 1). In these reduced units, the edge-lengths, particle volumes,  approximate order-disorder transition pressure (ODP)  and mesophase-crystal transition pressure (MCP) for all the particles studied are shown in Table \ref{tbl:shapedata}. The volume fraction $\phi$ of the system is just the ratio of the volume occupied by the polyhedral particles to the total volume 
of the simulation box.

\par Each pressure step of the expansion/compression run involved a total of $3 \times 10^6$  MC cycles (as defined below) for both equilibration and production. 
\begin{table}[h]
\small
  \caption{\ Reference data for the pure components studied. ODP describes order-disorder transition pressure which in most cases is an isotropic-mesophase transition. MCP denotes the mesophase-crystal transition pressure. The references used for the data are listed alongside the shape.}
  \label{tbl:shapedata}
  \begin{tabular*}{0.5\textwidth}{@{\extracolsep{\fill}}lllll}
    \hline
    Shape & Edge-length & Volume & ODP & MCP \\
    \hline
    Cube\cite{Agarwal2011}                 & 1  &  1 & 6.3 & 8\\
      Cuboctahedron\cite{Agarwal2012}        & 0.707 &  0.833 & 7.1 & 14 \\
    Sphere\cite{Vega2007} & 1 (diameter) & 0.524  & 11.5 & -\\
  Truncated Octahedron\cite{Agarwal2011}   & 0.354  &  0.5 & 14& 28\\
         \hline
  \end{tabular*}
\end{table}
 Each MC cycle consisted on average of N translational, N rotational, N/10 flip, N/20 swap and 2 volume move attempts. Flip moves attempt to rotate a chosen particle to a random orientation in the 
plane perpendicular to its present orientation. Swap moves involved picking randomly two particles, one of each species, and attempting to swap their positions as well as orientations (since such moves are more likely to succeed with swapped orientations). Swap moves are essential to speed-up the equilibration process by allowing particles
 to move arbitrarily far from their original positions  circumventing the slow diffusion associated with dense phases. Swap moves, however, can only be effective if the particle shape and size of the two components is similar and densities are  moderate to allow some wiggle room around a lattice site to accommodate different particle shapes. At pressures where the system is ordered (as obtained in trial runs), the volume moves were allowed to be triclinic \cite{Agarwal2011}. The size of the move perturbations was adjusted so as to get an acceptance probability of  0.4, 0.4 and 0.2 for the translation, rotation and volume moves, respectively. Although the size of the pressure steps was not fixed, a typical value of $ \Delta P^*$ near the phase transition was approximately 0.8, 1.3, 0.8 and 1.6 for CS, CTO, COC and COTO mixtures respectively. All trial moves are accepted according to the Metropolis criterion \cite{Metropolis1953} (which for hard-core interactions requires the absence of overlaps, 
checked via the separating axes theorem\cite{Golshtein1996} for any two polyhedra or  Arvo's algorithm for the cube-sphere case \cite{Arvo1990}).
\par Interfacial simulations where two phases and the intervening interfaces are present in the same box were used to estimate coexistence conditions. To facilitate the formation of distinct bulk regions, the box was at least four times more elongated along one direction that the others, so that the interfaces would form perpendicular to that axis. Once interfaces form, the longitudinal pressure (i.e., the one acting on a plane parallel to the interfaces) provides the proper estimation of the coexistence pressure since the transverse pressure contains a contribution from the surface tension \cite{Frenkel2013}. Thus, in such cases we chose the transverse box length (which sets the  dimensions of the interface) to be commensurate with the lattice spacing for the individual phases at the given pressure, only sparingly allowing changes in transverse dimensions to help relieve any build-up of stresses. Especial care must be taken when dealing with interfacial simulations involving two solid phases where the box cross section (perpendicular to the long axes) must be chosen so that it can properly accommodate the unit cells of the distinct crystal lattices. Unfortunately, this is hard to achieve over a wide range of pressures. Hence, although  triclinic volume moves  did allow for box deformations that can rotate the lattices to alleviate internal stresses, our results in such regions are expected to have larger errors than at lower pressures. 
\par Unless otherwise indicated,  interfacial simulations were performed for an equimolar global composition as it was expected that, if phase separation occurs, the compositions of the coexisting phases would be relatively symmetric (on account of the particle-size similarity) and hence lead to similar amounts of the two phases. To estimate the equilibrium bulk densities of the two phases in an interfacial simulation, we used density profiles along the z-axis to mark the bulk regions. 

\par We calculated the $Q_4$ and $Q_6$ bond-order orientational parameters \cite{SteinhardtNelson1983} to probe and monitor translational order. 
These parameters are defined as:
\begin{equation}
 Q_l =  \frac{4\pi}{2l+1} \left[ \sum_{-l}^{+l} |\bar{Q}_{lm} ({\bf r})| ^2 \right] ^{\frac{1}{2}}
\end{equation}
where $ \bar{Q}_{lm}({\bf r})$ is given by
\begin{equation}
 \bar{Q}_{lm}({\bf r} )= \frac{1}{N_b} \sum_{bonds} Y_{lm} ({\bf r})
\end{equation}
where $ Y_{lm}({\bf r})$ are spherical harmonics for the position vector $\bf r$.
Although the values of these order parameters are sensitive to the crystal structure, $Q_6$ is generically a good descriptor of crystallinity, since its value
increases monotonically with order. The value of the $Q_4$ order parameter gives additional information about the type of crystalline structure present in the system; 
i.e., larger values are associated with cubic symmetry. 
\par
\begin{figure}
 \centering
   \includegraphics[height=8.5cm]{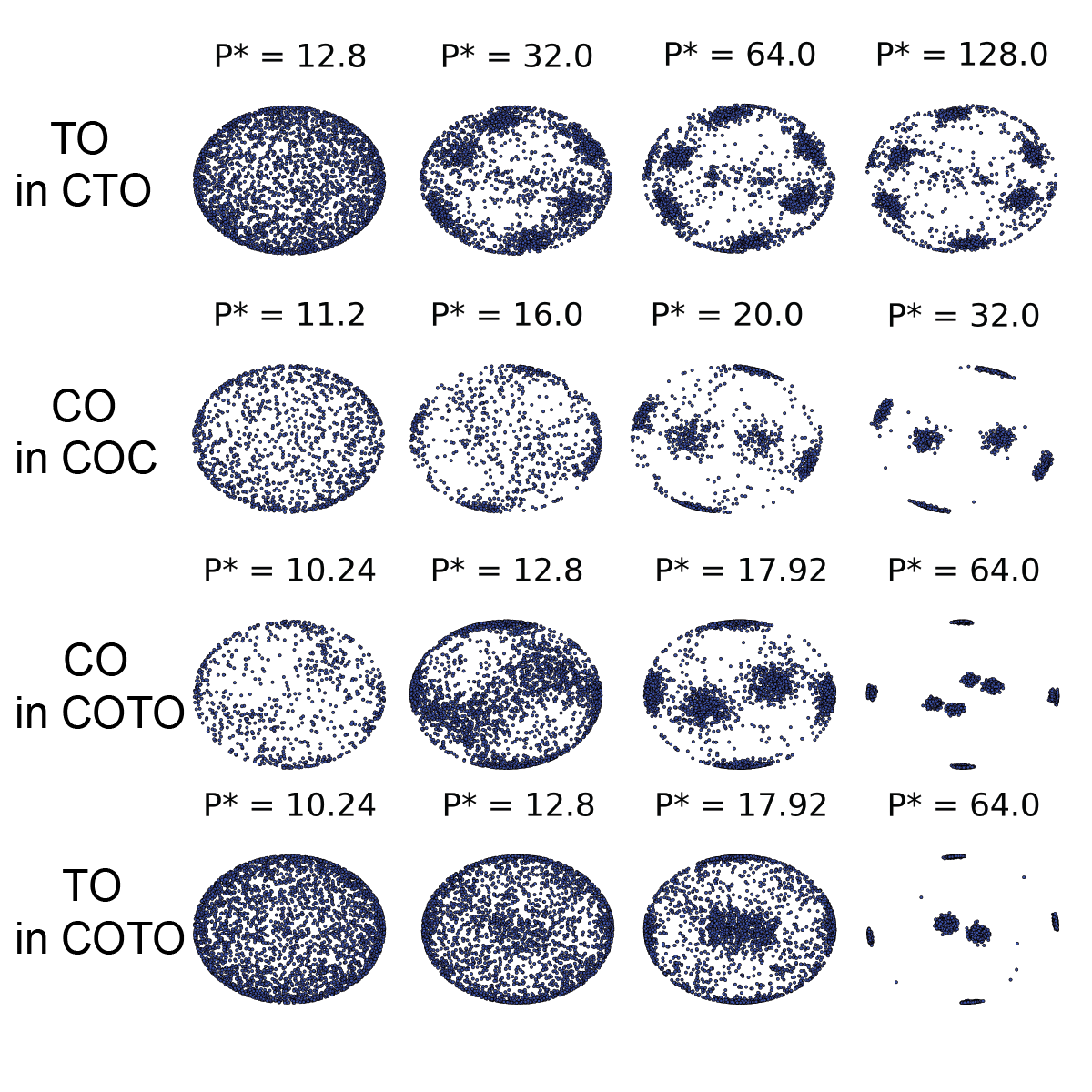}
   \caption{Representative orientational scatterplots used for determining the orientational order. From top a) Truncated Octahedra (TO) in CTO mixture, b) cuboctahedra (CO) in COC mixture, c) CO in COTO mixture, d) TO in COTO mixture. For pressures above the phase separation, the scatterplots show the orientational order for the particles in the ordered phase that is rich in the named component  only.}
  \label{fgr:scatter}
 \end{figure}

 To determine the orientational order in the system, we calculated the $P_4$ cubatic order parameter\cite{John2008} which gives information about cubic-like orientational order. To mark the boundary between the rotator mesophase and the crystalline phase, we used a combination of the $P_4$ order parameter and the orientational scatterplots. Since such a transition is continuous, we set  the upper limiting $P_4$ value of the rotationally disordered phase to be 0.3 (note that the maximum value of $P_4$ for perfect cubic order is  0.583). The scatterplots were generated by plotting the orientational axes of each the particles on a unit sphere. In these scatterplots, a mostly uniform distribution of points signals the absence of orientational order, which along with the presence of translational order (estimated through the $Q_6$ order parameter) identifies a rotator mesophase. A patchy pattern gives the signature of a particular orientational-order symmetry, which is absent in a rotator mesophase. A set of representative scatterplots for the systems studied here are shown in Figure \ref{fgr:scatter}.

\section{Results}
\subsection{Cubes + Spheres (CS) Mixture}

As previously indicated, this system represents  a base case of components with widely different shapes and monodisperse phase behavior. Cubes exhibit a mesophase\cite{John2005} for a small range of volume fractions (0.51-0.54) which separates isotropic and cubic crystalline phases\cite{Agarwal2011}. There is disagreement  in the  literature \cite{Smallenburg2012}\cite{Frenkel2012} about the precise nature of such a mesophase, which  hinges on the criterion adopted for classifying a phase as being  crystal-like or liquid-crystal-like. For the range of volume fractions specified above, while the {\it average} particle positions are crystal-like (despite a high vacancy content), the {\it variance} of the position fluctuations and particle mobilities are liquid-like \cite{Agarwal2012}. Since the current taxonomy is not completely satisfactory, for concreteness we will henceforth refer to this mesophase as cubatic. Spheres on the other hand, show a single phase transition from isotropic to FCC crystal. While spheres have full rotational symmetry, cubes do not and possess a relatively large   asphericity $\gamma$ (= ratio of circumradius to inradius = 1.732).  In our simulations, the side of cubic particle was set to be equal to the diameter of the spherical particle. 

\par Extensive MC runs for the equimolar mixture exhibit an isotropic, fully mixed phase, up to P* $ \approx$ 8, close to the MCP for cubes. Thus as  phase separation ensues above P* $\approx$ 8, the cubes are already in a crystalline state, while spheres are still disordered. At a pressure P* $\approx$ 12, the spheres order, a value which is only slightly higher than the ODP for pure spheres ($P*\approx 11.54$). Note that generally, for any A+B mixture, the apparent ODP in the A-rich phase is expected to be near but slightly above the ODP of the pure A system because of the disordering effect of the B particles present; our observations are consistent with this expectation. 
%an example of a two-column figure
 \begin{figure*}
   \centering
   \includegraphics[height=11cm]{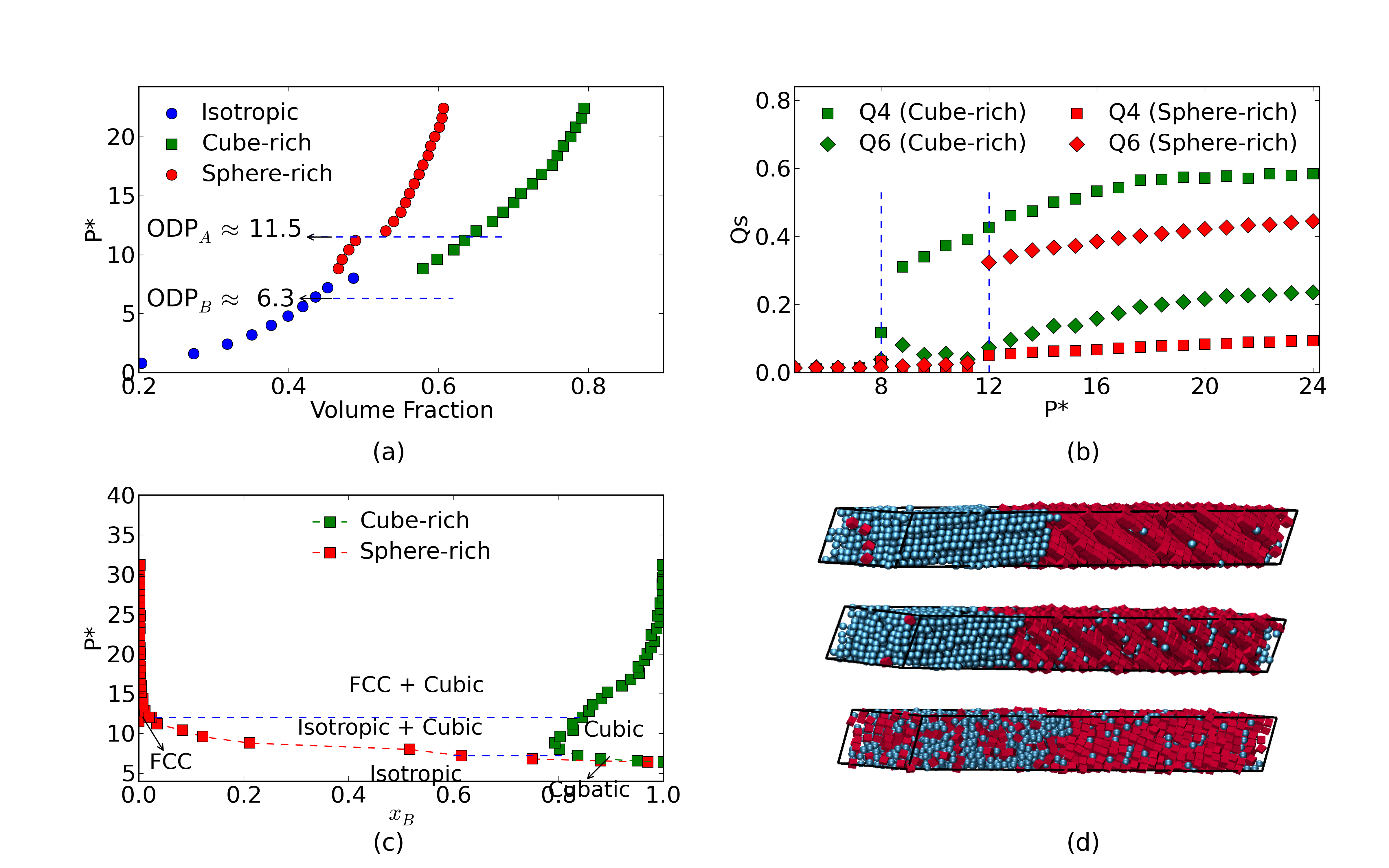}
   \caption{Summary of the results for the CS mixture. a) Equation of state with ODPs for the individual particles  shown for reference. Here ODP$_A$ is the ODP of spheres while ODP$_B$ is  the ODP of cubes. b) Plot of $Q_4$ and $Q_6$ order parameters. c) Pressure vs. composition phase diagram with different regions marked with the order of the constituent phases. Here $x_B$ is the mole fraction of cubes. d) Snapshots from the interfacial simulations of 3 representative regimes of the phase diagram (spheres are cyan and cubes are dark red). Bottom: cubic+isotropic phases  at P* $\approx$ 8. Middle:  Same phases  at P* $\approx$ 11. Top:  Cubic + FCC phases at P* $\approx$ 15.}
   \label{fgr:csplots}
 \end{figure*}

\par 
We demarcate the putative cubatic region by a combination of  increased $P_4$ order parameter and high translational mobility. We estimate the latter by calculating the mean square displacement (in a pseudo-dynamic setup) during compression runs at different pressures. A high value of translational mobility coefficient\cite{Agarwal2011} marks a region of cubatic phase. Mapping out the region of cubatic behavior was not possible for the  $50 \%$ global-composition system as the cube-rich phase formed by phase separation at even the lowest pressure  already had the cubes in a cubic lattice (with $\phi \approx 0.59$). We therefore performed additional simulations at $60\%$ to $90\%$ number composition of cubes at pressures just below the phase separation pressure. This helped us better define the isotropic-cubic and isotropic-cubatic boundaries in the phase diagram. 
\par  The approximate  Pressure vs. Composition phase diagram  is shown in Figure \ref{fgr:csplots}.  Note that even the ordered phases of each of the individual species allow a certain extent  of mixing with the other species. Further,   the cube-rich phase  solvates significantly more spheres  than the sphere-rich FCC phase solvates  cubes. This asymmetric solvation capacity seems to be primarily  the result of both our choice of particle size ratios which gives cubes a larger volume (and excluded volume) than that of spheres, and the geometry of the respective lattice spacings. Indeed, a sphere can readily replace a cube without overlap (if placed at the same center of mass) but not the other way round. Consequently, mixing entropy favors cubic phases where numerous spheres are allowed provided they do not take away the packing entropy gains associated with the overall cubic order. Conversely, cubes are only allowed in the FCC sphere-rich phase as 
long as they can appear as very dilute localized defects that do not compromise the overall structural order. 
\par In the bigger picture, this mixture can be seen to represent a `base case', of particles with very distinct shapes and incompatible lattice structures. In this scenario, although there is certain amount of mixing at intermediate concentrations, the ordered phases and the transition pressures remain quite similar to those of the pure components. This is expected since the two shapes are different enough that the mixing entropy to be gained from their intermingling is outweighed by the packing entropy lost due to the incompatibility of lattices.

\subsection{Cubes + Truncated Octahedra (CTO) Mixture}
This mixture represents a case of two space-filling polyhedra, which have incompatible crystal structures. While cubes form a cubic lattice in their ordered state, truncated octahedra (TOs) form a BCC tessellation. Further, pure TOs exhibit a rotator mesophase as an intermediate between the liquid and the ordered crystal phase. Figure 5 shows the main results for this system with Fig. 3 also showing some relevant scatter plots (in the top row). 
\par Upon extensive compression and expansion MC runs, the CTO mixture is observed to form a uniform isotropic phase  at low volume fractions up to $P^*\sim 7.7$ where  two phases emerge: a cube-rich cubic crystal and a TO-rich isotropic phase. 
 \begin{figure*}
   \centering
   \includegraphics[height=11cm]{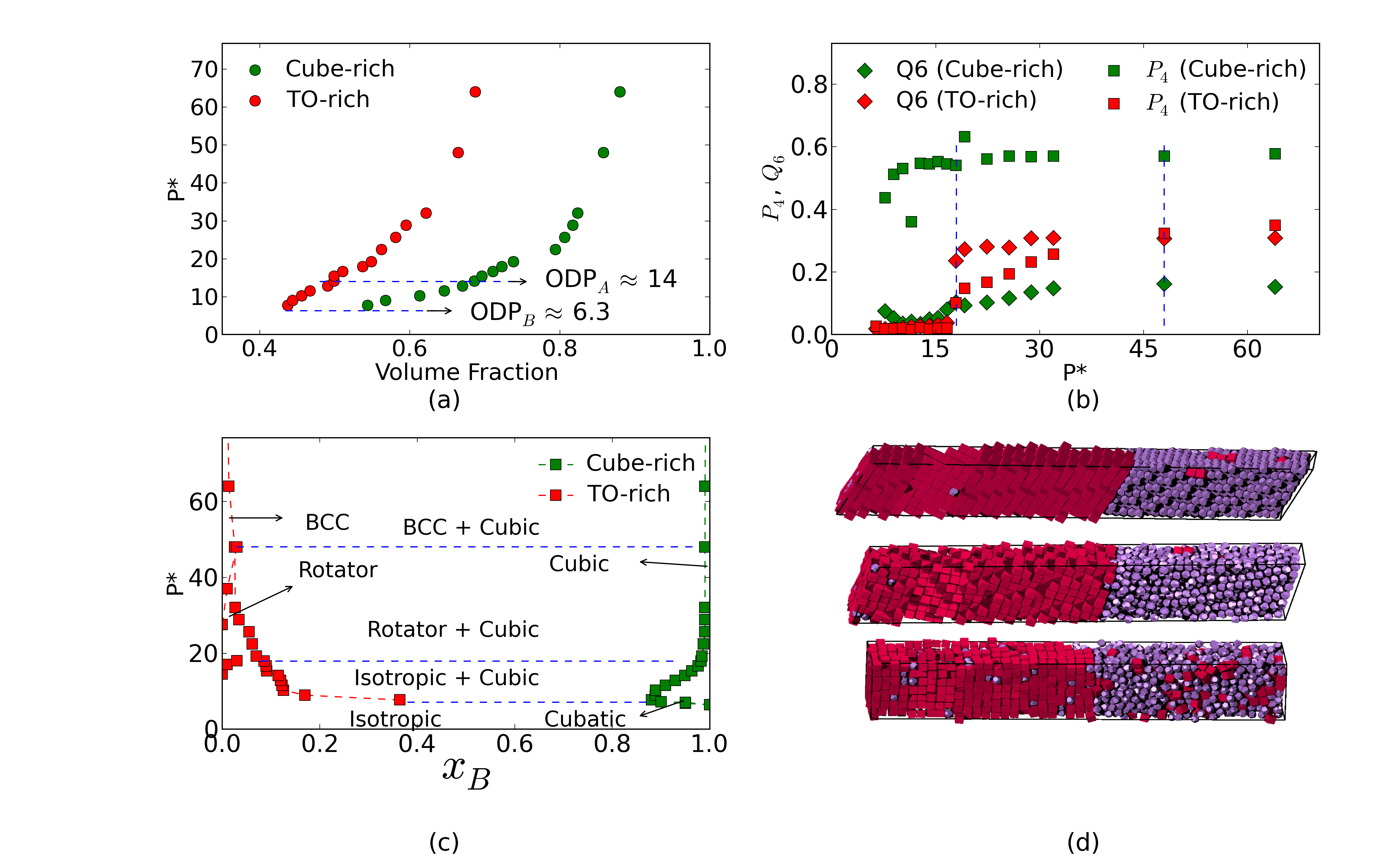}
   \caption{Summary of results for the CTO mixture. a) Equation of state with  ODPs for the individual particles marked for reference. Here ODP$_A$ is the ODP of TOs while ODP$_B$ is the ODP of cubes. b) Plot of $P_4$ and $Q_6$ order parameters. c) Pressure vs. composition phase diagram with different regions marked with the order of the constituent phases ($x_B$ is the mole fraction of cubes). d) Snapshots from the interfacial simulations of three representative regimes of the phase diagram (TOs are purple and cubes are dark red). Bottom: cubic+isotropic phases at P* $\approx$ 9. Middle:  cubic +rotator phases at P* $\approx$ 25. Top:  Cubic + BCC phases at P* $\approx$ 55.}
   \label{fgr:ctoplots}
 \end{figure*}

At P* $ \approx$ 18, the TO-rich phase  is seen to transition into a rotator mesophase, a value larger than that for  pure TOs which form a rotator mesophase at P* $\approx$ 14. This difference is expected since the presence of cubes in the TO-rich phase should make ordering more difficult and drive the ordering pressure upwards. The onset of the positional order in the rotator mesophase is pinpointed by calculation of $Q_6$ order parameter, which shows a sudden jump in $Q_6$ for the TO-rich phase. 
\par At   P* $ \approx$  50, the TO-rich phase achieves a crystalline BCC order (found through  $Q_4$ and $Q_6$ values) and orientational order (found through  $P_4$ values and orientational scatterplots), a pressure that is again larger than that for  pure TOs which gain crystalline BCC at around P* $ \approx$ 28.  To better define the cubatic region, we again perform independent runs away from the 50$ \%$ number composition (between $80\%$ and $95\%$) and assigned a cubatic character to systems exhibiting both high $P_4$ values and    a high particle translational mobility. 
\par As seen in Figure \ref{fgr:ctoplots}(c), neither the cube-rich nor the TO-rich crystal phases allow a significant concentration of  the other species. The saturation compositions tend to be symmetric; e.g., at intermediate pressures the cube-rich phase saturates with approximately 1\% TOs and the TO-rich phase saturates with about 1\% cubes. In a way, the CTO mixture shows less inter-species miscibility  than the CS mixture. This could be because: (1) a slightly larger size disparity reduces the entropic gain obtained from mixing entropy as the pressure increases, and (2) the space-filling TOs are less tolerant of impurities (cubes) than spheres (which are non-space filling). This question is revisited in Section \ref{sec:trends}.

\subsection{Cuboctahedra + Cubes (COC) Mixture}
The components of this mixture have some key similarities and differences with those of the CTO mixture. Like TOs in the CTO mixture, cuboctahedra (COs) in the COC mixture also exhibit a rotator mesophase over a significant range of volume fractions, before going into a crystalline phase. The crystalline
phase for pure COs, however, is a distorted cubic phase unlike  the BCC crystal exhibited by pure TOs. More importantly, the  two species in the COC mixture are quite close to each other in size, with a volume ratio of 0.83 (in the CTO mixture this ratio is 0.5).  Our main results for this system are shown in Fig. 6 (see also Fig. 3 for some sample scatterplots).
\par Compression and expansion MC runs show phase separation of the mixture at P* $\approx $ 9, just above the MCP for cubes. Hence, phase separation upon compression of the isotropic phase gives rise to a CO-rich isotropic phase and a cube-rich cubic crystal. At P* $\approx$ 14, the CO-rich phase transitions into a rotator mesophase, as evidenced by a sudden rise in the value of the $Q_6$ order parameter.  For pure COs, this transition happens around P* $\approx $ 7. As the pressure increases to around P* $\approx $ 20, the CO-rich phase undergoes another phase transition from rotator to crystalline phase, while the same transition happens at P* $\approx$ 14 for pure COs. 
\par
\begin{figure*}
   \centering
   \includegraphics[height=11cm]{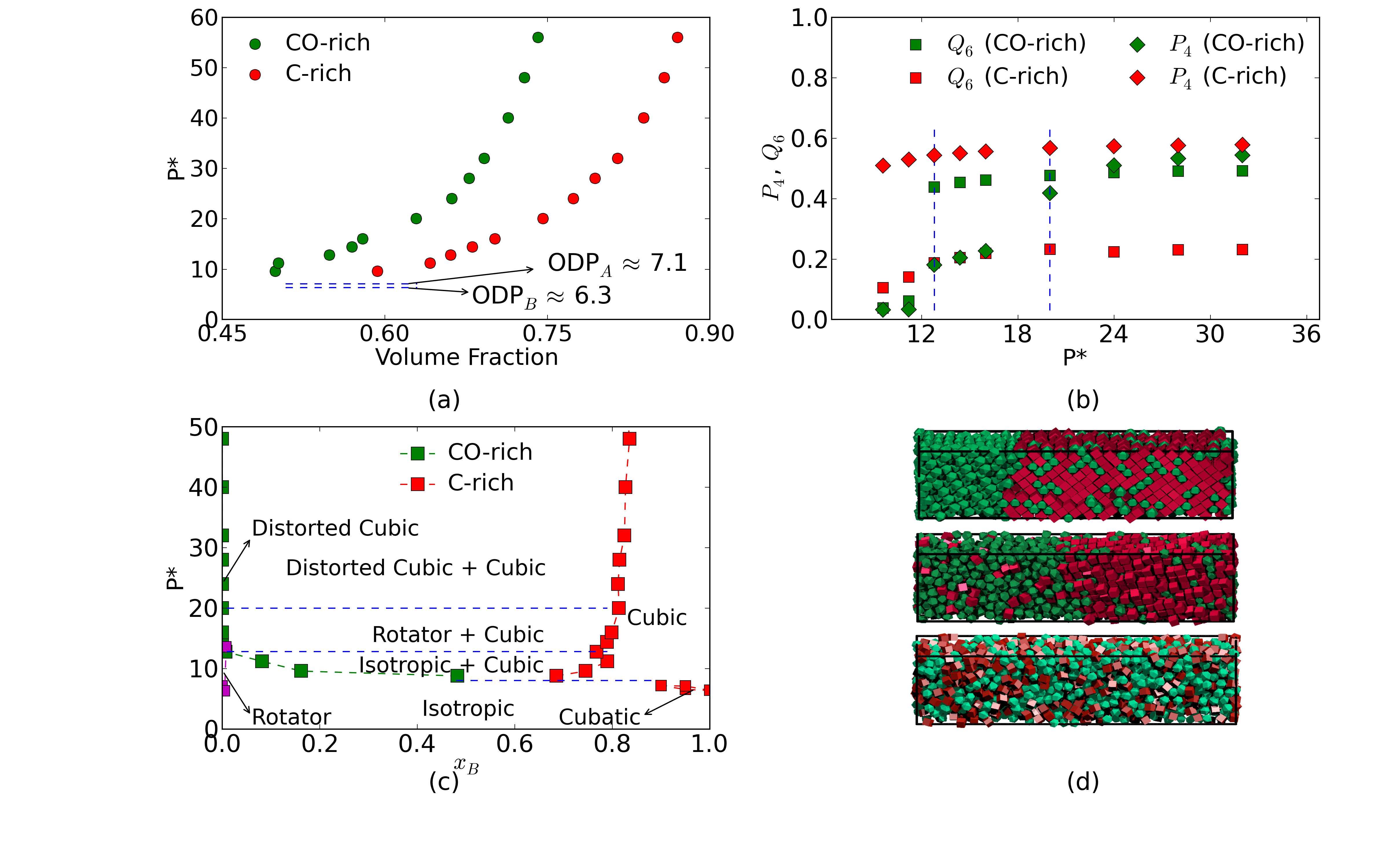}
   \caption{Summary of results for the COC mixture. a) Equation of state with  ODPs for the individual particles shown for reference. Here ODP$_A$ is the ODP of COs while ODP$_B$ is the ODP of cubes. b) Plot of $P_4$ and $Q_6$ order parameters. c) Pressure vs. composition phase diagram with different regions marked with the order of the constituent phases ($x_B$ is mole fraction of cubes). d) Simulation snapshots from  three representative regimes of the phase diagram (COs are green and cubes are  dark red) . Bottom: isotropic mixture phase at P* $\approx$ 6. Middle:  cubic + isotropic phases at P* $\approx$ 10. Top:  Cubic + distorted cubic phases at P* $\approx$ 25.}
   \label{fgr:cocplots}
 \end{figure*}

 An important difference observed in the Pressure vs. Composition phase diagram for the COC mixture (Fig. \ref{fgr:cocplots} (c)) as compared to the earlier mixtures is the relatively large saturation concentration of COs in the cube-rich phase (reaching about 30\%) even at high pressures. This enhanced mixing may partially arises due to  the COs gaining packing entropy by following orientations compatible with those of neighboring cubes. Further, near-equal volumes of the two phases imply that the packing entropy cost of allowing a CO-impurity in cubes is minimal. As in the previous systems, the low solubility of cubes in CO-rich phase is related to the bigger size of the cubes that makes them hard to be accommodated as guests in the CO-rich phase at high densities. See further analysis in Section \ref{sec:trends}.
\subsection{Cuboctahedra + Truncated Octahedra (COTO) mixture }
The COTO mixture represents a case where the two shapes involved are similar in terms of their individual phase behavior. Further, they are neighbors in shape evolution (during polyol process). While truncated octahedra form a BCC tessellation, cuboctahedra arrange in a distorted cubic lattice which is non-space-filling. However, both shapes show a rotator mesophase for a wide range of volume fractions. 

\par While one could have expected these two shapes to be the most miscible of the mixtures considered here, our simulation results shown in Fig. 7 (and in the relevant scatterplots of Fig. 3) provide evidence to the contrary. This is primarily because although the similarity between shape and individual phase behavior could promotes miscibility, the size ratios adopted here (based on the polyol process) make cuboctahedra significantly bigger than truncated octahedra. Both the volume ratio (1.6) and the ratio of the radii of their circumspheres ($\approx 1.25$), translate into a twofold difference in the componentsÍ ODP values  (see Table \ref{tbl:shapedata}) and a scant miscibility.
\par MC compression runs, which start from an isotropic liquid state fail to phase-separate and instead get kinetically trapped into a disordered state. It appears that at pressures where the mixture would prefer phase-separating, the mixtures is already too dense for a well-mixed system to demix and phase-separate. However, expansion runs started from two-phase ordered configuration (with individual shapes ordered in their corresponding crystal lattices) is observed to be stable in a two-phase state down to very low volume fractions. We thus use expansion runs in this case to estimate thermodynamic phase behavior. The fact that the compression and expansion runs yield different results at high pressures signals lack of ergodicity (assumed to be more severe for the compression process). While unphysical MC moves like particle-swaps help overcome diffusion barriers to compositional equilibration,  more elaborated moves may be needed to speed up structural equilibration (and the nucleation of translational order).

\par Below P* $\approx$ 30 (which is slightly above the MCP for TOs), the TO-rich phase goes into a rotator phase while the CO-rich phase is still ordered in a distorted cubic lattice. Below P* $\approx$ 18, again slightly above the MCP for COs, the CO-rich phase too becomes  a rotator phase. In a narrow region between P* = 13.5 and P* = 11.5, we see the TO-rich phase in an isotropic phase while the CO-rich phase remains as a rotator mesophase. Below P* = 11.5, the two phases mix to form a single isotropic phase.
\par As far as inter-species miscibility, in the small region (between P* of 11.5 and 13.5) where some miscibility occurs, COs are seen to be more miscible in the TO-rich phase than the other way around. This is expected since in that region the TO-rich phase is largely isotropic which makes the entropic cost of introducing a CO impurity in the TO-rich phase minimal. On the other hand, the CO-rich phase is positionally ordered and loses some packing entropy  by hosting  TO-impurities. This mixture demonstrates the fact that along with shape and individual phase behavior, the relative size of the particles can have a dominant effect on the mixture phase behavior.  \begin{figure*}
   \centering
   \includegraphics[height=11cm]{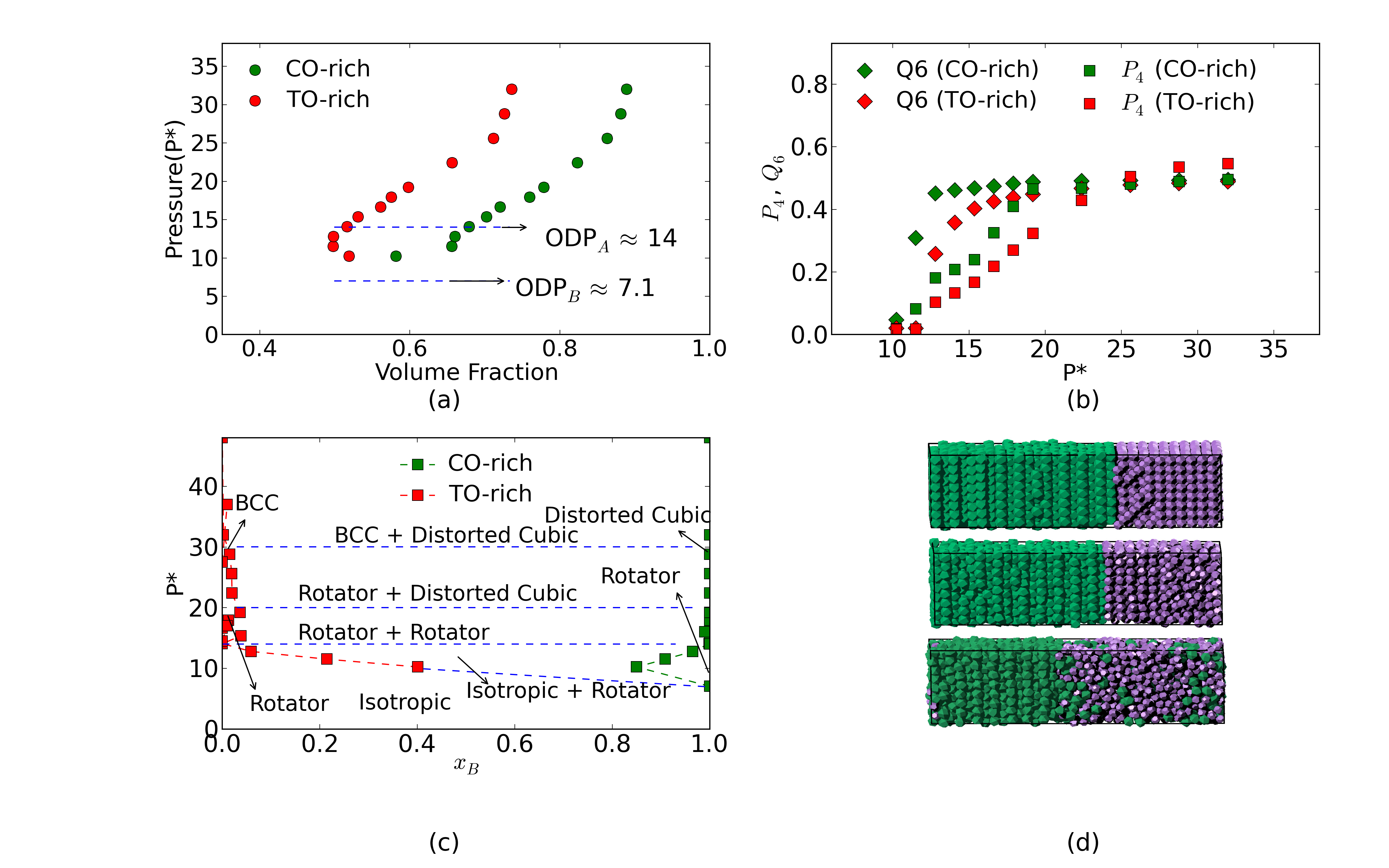}
   \caption{Summary of results for the COTO mixture. a) Equation of state showing the  ODPs of the individual particles  for reference. Here ODP$_A$ is the ODP of TOs while ODP$_B$ is the ODP of COs. b) Plot of $P_4$ and $Q_6$ order parameters. c) Pressure vs. composition phase diagram with different regions marked with the order of the constituent phases ($x_B$ is the mole fraction of COs). d) Simulation snapshots of three representative regimes of the phase diagram (COs are green and TOs are dark purple) . Bottom: rotator + isotropic mixture phase at P* $\approx$ 12. Middle:  distorted cubic + rotator mixture phase at P* $\approx$ 20. Top:  distorted cubic  + BCC phase at P* $\approx$ 32.}
   \label{fgr:cotoplots}
 \end{figure*}
\section{\label{sec:trends}{Discussion of general trends and comparison to other systems}}
\begin{figure}
\centering
\includegraphics[height=4cm]{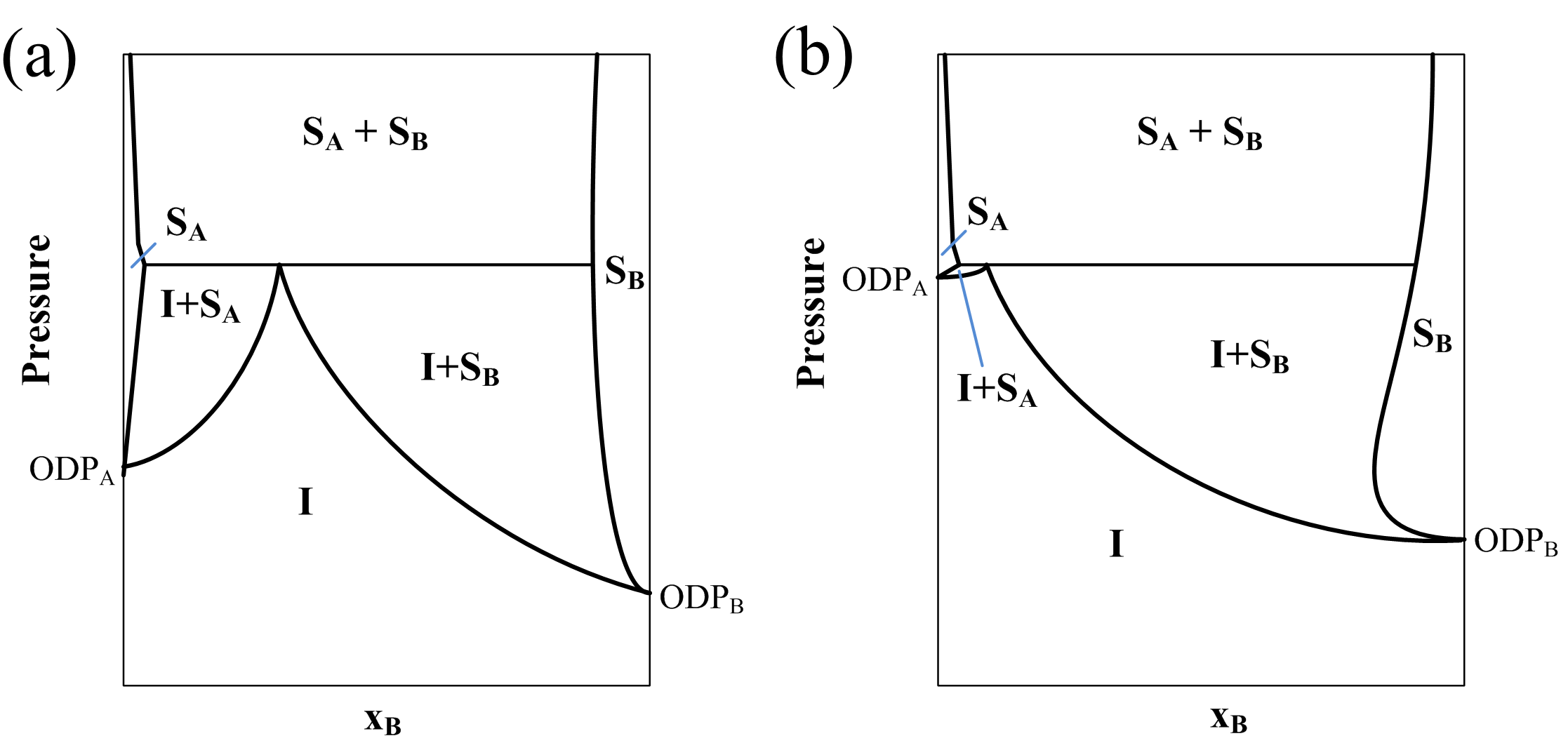}
 \caption{Qualitative sketches of pressure-composition phase diagrams for a binary mixture of hard spheres of diameter ratio $\alpha$=0.85 (a), and for the CS, CTO, COC, and COTO mixtures (b). I = Isotropic, $S_A$, $S_B$ = rich A and rich-B solids respectively, with A(B) being the smaller (larger) component}
\label{fgr:schematic}
\end{figure}

Systems CS, CTO, and COC constitute similar mixtures of a large component B, cubes, with a smaller component A that has moderate asphericity ($\gamma<$1.5) and high rotational symmetry. The COTO mixture is slightly different but can also be cast as a A+B mixture with the TO and CO corresponding to species A and B. Overall these mixtures seem to exhibit a eutectic type of phase behavior, akin to that formed by binary A+B mixtures of hard spheres whose size ratio $\alpha$ (=diameter of A/diameter of B) is different enough (say $\alpha \approx $0.85) so that at high pressures they tend to phase separate into two incompatible FCC lattices\cite{Kranendonk1991}\cite{Punnathanam2006}. Figure \ref{fgr:schematic} shows a qualitative diagram for such a case along with one that encapsulates the behavior observed in the CS, CTO, and COC systems. Note that our simulations were unable to resolve neither the position of the eutectic point (which in all cases seems to be near zero cube composition) nor the very small isotropic+solid A phase coexistence region. Despite the quantitative disparities in the diagram proportions and the differences in the lattice symmetry of the ordered phases, the physics underlying the phase diagrams (a) and (b) in Figure \ref{fgr:schematic} is the same:  just like in our systems, in the mixture of hard spheres phase separation at high pressure is driven by the maximization of packing entropy that results from forming two distinct efficiently-packing solid phases. At intermediate pressures where (only) the pure B system would solidify (as it has the lower ODP), a B-rich ordered phase must form which hence coexists with an isotropic phase within a large two-phase region. In this context, the fact that in our systems one or the two components have flat facets does not fundamentally change the overall picture, that merely leads to ordered structures that depart from the ones favored by spheres. Also, the fact that our systems exhibit mesophases (that precede the perfect crystal at high pressures) does not add anything fundamentally new in the character of these diagrams since all mesophase-crystal phase transitions are continuous and hence such mesophases simply occupy the lower-pressure portion of the A-rich or B-rich solid regions. 

\par
\begin{figure}
\centering
\includegraphics[height=7cm]{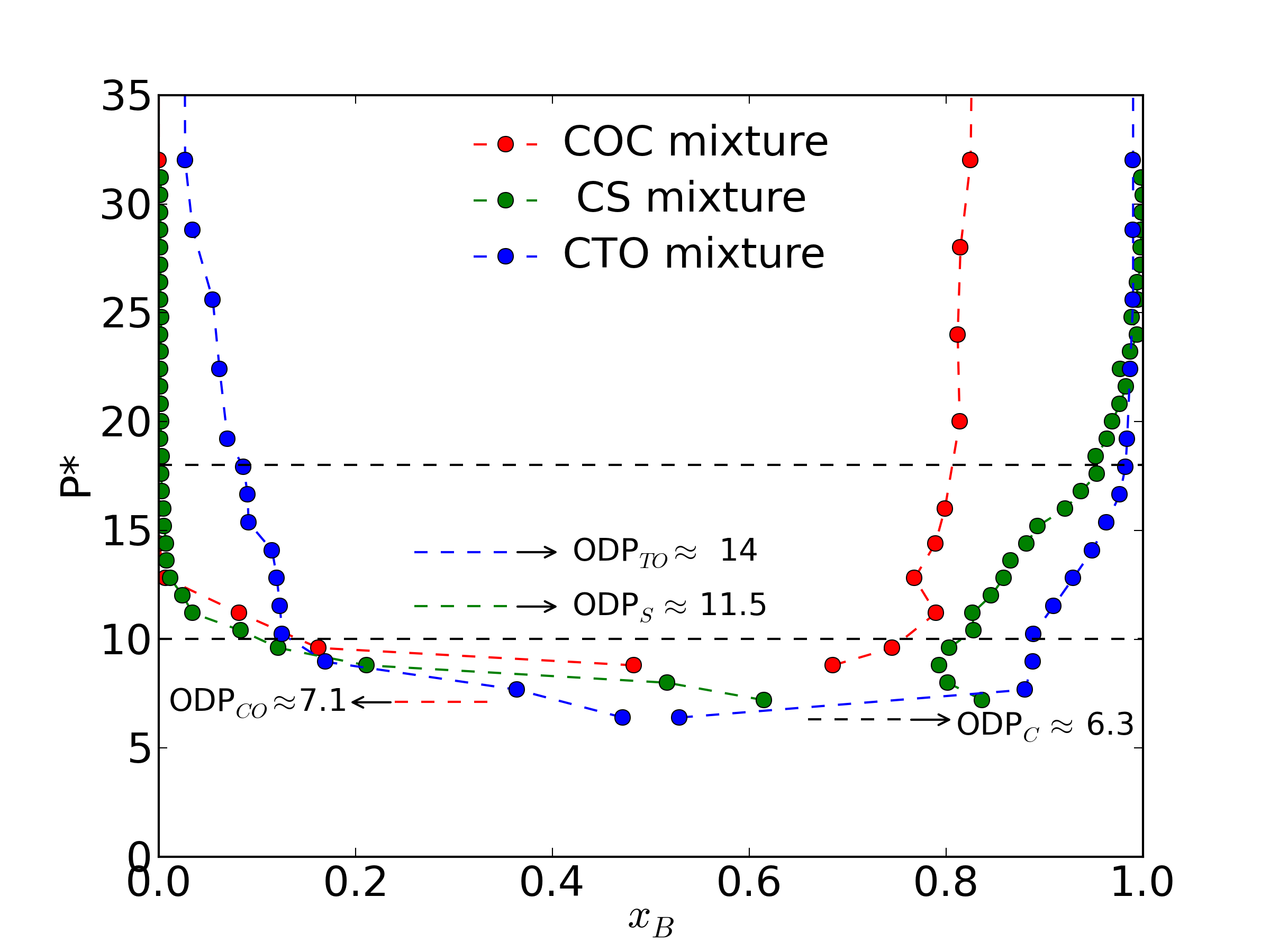}
 \caption{Pressure-composition phase diagrams for the CS, CTO  and COC mixtures overlaid on top of each other. For simplicity,  only the main two-phase envelopes are shown, leaving the boundaries connecting to the pure-species ODPs untraced . Two particular P* values, P* = 10 and P* = 18 are marked for reference. $x_B$ is the mole fraction of cubes.}
\label{fgr:overlay}
\end{figure}

 Another general trend relevant to the CS, CTO, and COC mixtures (excluding now the COTO mixture) relates to the correlation between component relative sizes and inter-phase solubility (i.e., how much of A can be dissolved in a B-rich phase and vice versa). Here cubes (the common component) can be taken as the reference whose size sets not only a reference unit length but also a common baseline to directly compare pressure values. In such a case, the pure components rank in order of decreasing size ( $\sim $ volumes) as cubes, COs, spheres, and TOs, which is also the order of increasing ODPs. This shows that the ODPs are more strongly affected by differences in particle size (than in particle shape). To examine how the saturation solubilities compare for the same pressure across mixtures, we ovelay the pressure-composition diagram for these 3 mixtures in the same plot (Figure \ref{fgr:overlay}) and consider (for concreteness) two cases: P*=10 for the isotropic-cubic coexistence region and P*=18 for the rotator-cubic coexistence region (selecting other pressures would give similar results). 

For a fixed P* we will denote as the most compressed A-rich phase (among the 3 mixtures) the one that has the lowest ODP$_A$. (i.e., the degree of compression is relative to ODP$_A$). This is rooted on the fact that at the ODP, the isotropic phase for all systems considered has a similar packing fraction and so one can assume that at the ODP all A-rich (or B-rich) systems are comparably dense. We see in Figure 9 that at a given pressure: (1) Cubes tend to dissolve more into the A-rich phase (whether it is isotropic or rotator) for the mixture that has the highest ODP$_A$, namely, where the A-rich phase is less compressed (at the given P*), and (2) conversely, the A component dissolves more into the cubic phase for the mixture that has the lowest ODP$_A$, namely, for the A-rich phase that is more compressed (at the given P*). Both of these trends are consistent with the idea that a component that is in a more compressed phase has a higher chemical potential (and a higher tendency to escape) and would then have a higher relative proclivity to transfer to the other phase so that chemical potentials can be equalized. For trend (1) cubes in the cubic phase are always equally compressed (for a given P*) but will transfer to and populate more a coexistence phase that is less compressed and hence more hospitable to guests; for trend (2) the A component experiences different states of compression in the A-rich phases across mixtures and hence escapes to different extents into the coexistence cubic phase (that is always equally compressed at a given P*). Because our systems are purely entropic, a higher (lower) relative degree of compression or chemical potential translates to essentially a lower (higher) entropy. Of course, these trends are only approximate and expected to hold provided that, among different systems, the ODPs are sufficiently different but one is still comparing at conditions where similar phases coexist. Note that other pure-system attributes (besides ODPs) could be used to try to correlate the observed miscibility trends; e.g.,the particle volume which, as Table 1 shows, consistently decreases as ODP increases and hence it is an equally good descriptor. Further, some particular function of ODP or particle volume may prove to be better at correlating those trends in a more quantitative way. Our limited sample of mixtures only allows us to point out qualitative correlations and to conjecture that relative values of ODPs,  landmarks of  pure component phase behavior that also capture particle size disparities, likely provide more robust clues of mixture phase behavior than any single geometrical feature of a particle shape. Of course, the precise particle shapes must also play a role in reinforcing or opposing these broad trends. The COTO mixture does not allow a direct comparison with the other mixtures for a fixed pressure but we expect that similar principles should apply on how the relative compression states of the phases affect the saturation solubility of the guest component. 
\section{Conclusions and a roadmap for equimolar phase behavior}

Towards the goal of identifying key factors that govern  the self-assembly of mixtures of polyhedral nanoparticles, we studied here the phase diagrams of four representative binary mixtures of hard convex polyhedra, using as components particle shapes that are readily accessible via well-established synthesis methods. In particular, we examined how the properties of the individual components affect the interplay between mixing and packing entropy which ultimately determines the types of phases formed and the extent of inter-particle mixing in such phases.
\par We find that, while the pure-component phase behavior is determined by  the rotational symmetry and asphericity of the particle shape, the binary-component phase behavior depends on both the pure-component phase behavior and on the  relative size ratio of the components, which in turn determines their  relative difference in ODPs.  For instance, in the COC mixture a combination of a similarity of size ratio and  pure-component crystal lattices makes COs particularly more miscible in the cube-rich ordered phase.  
\par  Casting our systems as  A+B mixtures where B is the component with the largest size and hence smaller ODP, we expectedly find that the relative extent of miscibility of the two shapes (miscibility of  A in the B-rich phase vs. that of  B in the A-rich phase) depends on the relative ODP values. In particular, if $\Delta$ ODP = ODP$_A$ $-$ $ODP_B$ with ODP$_A >$ ODP$_B$, then at any particular pressure where phase separation occurs, the larger the $\Delta$ ODP the lower the solubility of A in the B-rich ordered phase and the higher the solubility of B in the A-rich ordered phase
\par
\begin{figure*}
   \centering
   \includegraphics[height=11cm]{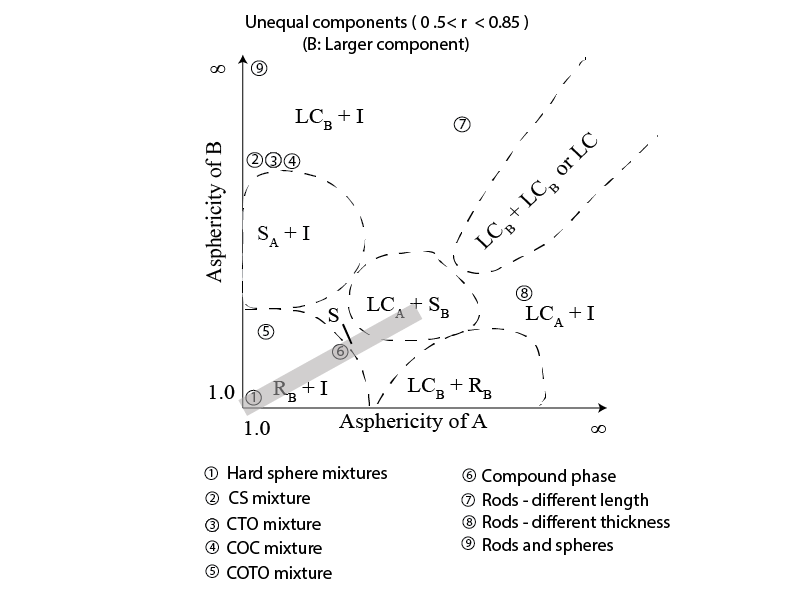}
   \caption{Tentative sketch of road map of phases at the ODPEM (the ODP of an equimolar binary mixture) of A+B hard particles for components with volume ratio $0.5 < $r$ < 0.85$.  I = isotropic, LC= liquid crystal, R = rotator solid, S = crystal solid; a subscript denotes the majority component in the phase. Gray region in (a) corresponds to single compound solids. Numbered circles denote selected states studied in the literature (cases 2 through 5 are from this work)  as follows: 1 = Ref. \cite{Kranendonk1991}, 6 = Ref. \cite{Khadilkar2012}, 7 = Ref. \cite{Lekkerkerker1983} and \cite{Escobedo2003}, 8 = Ref. \cite{Purdy2005}, and 9= Ref. \cite{Vliegenthart1999}}
      \label{fgr:globalphased}
 \end{figure*}

 
We attempt now to sketch out a rough phase roadmap that identifies the phases formed at the ODP of an equimolar mixture of hard particles (henceforth denoted the ñODPEMî); i.e., the first single- or two-phase state involving at least one ordered phase that arises upon compression of an isotropic equimolar mixture of A+B. We only consider mixtures consisting of shapes that for any particular asphericity $\gamma$  exhibit (as pure components) one or more of the following mesophases or ordered states: rotator (R), solid crystal (S), and liquid crystal (LC). If a given particle forms multiple ordered states, it is assumed that a LC occurs at a lower packing fraction (and pressure) than an R phase, which in turn would occur at a lower packing fraction than an S phase. Very high $\gamma$  values are assumed to be accessible with prolate or oblate shapes only, which should lead to LC behavior and low ODPs.
 The tentative roadmap shown in Figure \ref{fgr:globalphased}  is based on observations from this work and those from selected previous studies on  binary mixtures of hard particles (marked by numbers in the plot), and is  restricted  to intermediate particle volume ratios $r$  between $0.5 <  $r$ < 0.85$, which is the range that  our mixtures fall into. 

\par The diagram of Figure \ref{fgr:globalphased} is guided by the following observations. If B is the largest component, then it is expected that for low to moderate $\gamma_A$, ODP$_A>$ODP$_B$ and hence at the ODPEM (slightly above ODP$_B$) a phase separated state should ensue comprising an ordered B-rich phase (R, S, or LC depending on the $\gamma_B$ value) and an isotropic phase; however, for very large $\gamma_A$ the pure A component would be expected to form a LC with ODP$_A<$ODP$_B$ in which case at the ODPEM an A-rich LC phase should coexist with a B-rich isotropic phase. Of course, a crossover behavior could exist between these small-$\gamma_A$ and large-$\gamma_A$ regimes, where two ordered phases coexist at the ODPEM.
 In  Figure \ref{fgr:globalphased} we mark only the ordered phases that could be formed, if any such phase will form at all. The alternate outcome would be the formation of some type of jammed state  without  a well-defined structural order. The shape and extent of each region are only qualitative and meant  to guide the eye. A secondary particle shape parameter besides $\gamma$ would be necessary to make more discriminative diagrams ( rotational symmetry would be a good candidate \cite{Agarwal2011}). Note that we  assumed that  as $\gamma$ approaches 1, particles have higher rotational symmetry and we ascribed to crystal phases of spheres a rotator character since any infinitesimal departure from $\gamma$ = 1 would lead to a rotational degree of freedom. 
\par Compound crystal phases are known to exist for a number of binary hard-core particles, like the {\em Laves} phases for  unequal hard spheres\cite{Filion2009} \cite{Hynninen2007}or polyhedra that form tessellating compounds\cite{Khadilkar2012}. However, these may not be the phases that arise at the ODPEM (i.e., at equimolar composition) or may arise for  components with $r < 0.5$ and so they would not be included in Figure \ref{fgr:globalphased}. Likewise, a single mixed LC phase would only be a possibility for very specific types of component shapes. These two scenarios (where a single S or LC phase forms at the ODPEM) do not seem to correlate strongly with sphericity (other than a loose tendency of components to have similar $\gamma$) and so are only included in Figure \ref{fgr:globalphased} (as a gray region) for completeness. 
\par Altogether, our observations  highlight the fact that although many factors such as relative size, asphericity, individual crystal lattices, and mesophase formation determine the phase behavior of a mixture, the trends in mutual miscibility are best captured by the  components asphericities and their relative ODPs. Towards designing novel nanoparticle superstructures with desired properties, this study  hence provides  some guiding principles about the phase behavior of the binary mixtures derived from the properties of the constituent shapes. While all systems studied here are relatively asymmetric in terms of size (and ODP) values, we are currently exploring more symmetric binary systems where our preliminary results have already revealed significant differences with some of the trends observed here. Also, while the systems studied here involved convex particles only, the use of concave particles, especially when paired with complementary-shaped convex partners, would open the door to much more complex phase behaviors.
 
\subsection*{Acknowledgements}
Work on mixtures of two polyhedral particles was supported by the U.S. National Science Foundation Grant No. CBET 1033349. Work associated with mixture containing hard spheres was supported by the U.S. Department of Energy, Office of Basic Energy Sciences, Division of Materials Sciences and Engineering under award Grant No. ER46517.

\footnotesize{
\bibliography{rsc} %your .bib file
\bibliographystyle{rsc} %the RSC's .bst file
}

\end{document}